\documentclass[reprint,
 superscriptaddress,
 amsmath,amssymb,
 aps,
pra,
]{revtex4-2}

\usepackage[utf8]{inputenc} 
\usepackage[T1]{fontenc}    
\usepackage{hyperref}       
\usepackage{url}            
\usepackage{booktabs}       
\usepackage{amsfonts}       
\usepackage{nicefrac}       
\usepackage{microtype}      
\usepackage{graphicx}
\usepackage{xcolor, etoolbox}
\usepackage{amsmath,amsfonts,amsthm} 
\usepackage{mathtools}
\usepackage{ragged2e}
\usepackage{braket}      
\usepackage[english]{babel} 
\usepackage{enumitem}
\usepackage{xspace}
\usepackage{dsfont}
\usepackage{bm}
\usepackage{bbm}
\usepackage{dcolumn}
\usepackage{mathrsfs}

\hypersetup{
    colorlinks,
    linkcolor=blue,
    citecolor=blue,
    urlcolor=blue,
    breaklinks=true 
}

\begin{document}

\title{General class of continuous variable entanglement criteria}

\author{Martin G\"{a}rttner}
    \email{martin.gaerttner@uni-jena.de}
    \affiliation{Institut f\"{u}r Theoretische Physik, Universit\"{a}t Heidelberg, Philosophenweg 16, 69120 Heidelberg, Germany}
    \affiliation{Physikalisches Institut, Universit\"{a}t Heidelberg, Im Neuenheimer Feld 226, 69120 Heidelberg, Germany}
    \affiliation{Kirchhoff-Institut f\"{u}r Physik, Universit\"{a}t Heidelberg, Im Neuenheimer Feld 227, 69120 Heidelberg, Germany}
    \affiliation{Institute of Condensed Matter Theory and Optics, Friedrich-Schiller-University Jena, Max-Wien-Platz 1, 07743 Jena, Germany}
\author{Tobias Haas}
    \email{tobias.haas@ulb.be}
    \affiliation{Centre for Quantum Information and Communication, École polytechnique de Bruxelles, CP 165, Université libre de Bruxelles, 1050 Brussels, Belgium}
\author{Johannes Noll}
    \email{johannes.noll@stud.uni-heidelberg.de}
    \affiliation{Kirchhoff-Institut f\"{u}r Physik, Universit\"{a}t Heidelberg, Im Neuenheimer Feld 227, 69120 Heidelberg, Germany}

\begin{abstract}
We present a general class of entanglement criteria for continuous variable systems. Our criteria are based on the Husimi $Q$-distribution and allow for optimization over the set of all concave functions rendering them extremely general and versatile. We show that several entropic criteria and second moment criteria are obtained as special cases. Our criteria reveal entanglement of families of states undetected by any commonly used criteria and provide clear advantages under typical experimental constraints such as finite detector resolution and measurement statistics.
\end{abstract}

\maketitle

\textit{Introduction} --- 
Entanglement is an intrinsic quantum phenomenon and indispensable for quantum technologies. Among the many physical platforms used for quantum computing and simulation, systems described by continuous variables such as photonic systems \cite{Dong2008,Schneeloch2019,Asavanant2019, Mikkel2019,Qin2019,Moody2022} and cold quantum gases \cite{Gross2011,Strobel2014,Peise2015,Fadel2018,Kunkel2018,Lange2018,Chen2021,Viermann2022} are gaining importance. As entanglement is central to questions of, e.g., quantum thermalization \cite{Popescu2006,Kaufman2016}, information scrambling \cite{Gaerttner2017,Landsman2019}, quantum supremacy \cite{Jozsa2003}, metrology \cite{DAriano2001,Giovanetti2011} and quantum phase transitions \cite{Osborne2002}, efficient tools for its detection are crucial for the understanding of quantum phenomena.

The quest for entanglement criteria has a long history in the quantum optics literature, starting from early works by Duan, Giedke, Cirac, and Zoller (DGCZ) \cite{Duan2000}, Simon \cite{Simon2000}, and Mancini, Giovannetti, Vitali, and Tombesi (MGVT) \cite{Mancini2002, Giovannetti2003}, who formulated entanglement criteria based on the variances of measured field quadratures, which are necessary and sufficient for Gaussian states \cite{Braunstein2005b,Weedbrook2012,Serafini2017,Lami2018}. The intuitive reasoning underlying all of these criteria is that, for separable states, the fluctuations in pairs of non-local variables, like $\boldsymbol{X}_1 + \boldsymbol{X}_2$ and $\boldsymbol{P}_1 - \boldsymbol{P}_2$, are lower bounded by the uncertainty principle, while these bounds can be submerged for entangled states \cite{Horodecki2009}. Later, these criteria were refined by quantifying uncertainty through entropies of measured distributions of the field quadratures, i.e. marginals of the Wigner function, by Saboia, Toscano, and Walborn (STW) \cite{Walborn2009,Saboia2011,Tasca2013,Walborn2011} and others \cite{Schneeloch2018,Schneeloch2019}. Other methods for certifying continuous variable entanglement use modular variables \cite{Gneiting2011,Carvalho2012}, higher moments \cite{Shchukin2005,Haas2023}, specific algebras \cite{Agarwal2005} or information theoretic quantities \cite{Rodo2008,Gessner2017,Haas2021b}.

Here, we derive a general class of entanglement criteria based on the Husimi $Q$-distribution \cite{Husimi1940,Cartwright1976,Lee1995}. The key ingredient of our approach is that for the Husimi $Q$-distribution, in contrast to the Wigner distribution, the uncertainty principle can be stated in a strikingly general way, formalized by the Lieb-Solovay theorem \cite{Lieb2014,Schupp2022}. Combining this with the Peres-Horodecki criterion \cite{Peres1996,Horodecki1996,Horodecki2009} yields a separability bound for any concave function, generalizing our previous results on Husimi-$Q$-based witnesses \cite{Haas2022a}. We showcase the strength of these criteria by giving examples where they outperform previously known criteria and demonstrating advantages for experimental entanglement detection. An extended discussion is provided in \cite{PRA}.

\textit{Notation} --- We work in natural units $\hbar = 1$. We denote quantum operators by bold letters, e.g. $\boldsymbol{\rho}_{12}$, and vacuum expressions with a bar, e.g. $\bar{Q}$.

\textit{Phase space distributions} --- We consider a bipartition of a continuous variable quantum system described by operators $\boldsymbol{X}_j$ and $\boldsymbol{P}_j$ fulfilling bosonic commutation relations $[\boldsymbol{X}_j, \boldsymbol{P}_k] = i \delta_{j k} \mathds{1}$ with $j,k \in \{1,2 \}$ denoting the two subsystems. We allow for local rotations $\vartheta_j \in [0, 2 \pi)$ yielding rotated quadratures
\begin{equation}
    \begin{pmatrix}
    \boldsymbol{R}_j \\ \boldsymbol{S}_j
    \end{pmatrix} = 
    \begin{pmatrix}
    \cos \vartheta_j & \sin \vartheta_j \\
    - \sin \vartheta_j & \cos \vartheta_j
    \end{pmatrix}
    \begin{pmatrix}
    \boldsymbol{X}_j \\ \boldsymbol{P}_j
    \end{pmatrix}.
    \label{eq:RotatedVariablesDefinition}
\end{equation}
Their bosonic commutation relations induce sets of local coherent states $\ket{\alpha_j} = \boldsymbol{D} (\alpha_j) \ket{0_j} = e^{\alpha_j \boldsymbol{a}^{\dagger}_j - \alpha^{*}_j \boldsymbol{a}_j} \ket{0_j}$, where $\boldsymbol{D}(\alpha_j)$ denotes the canonical displacement operator acting on the vacuum state $\ket{0_j}$ \cite{Zhang1990,Radcliffe1971,Gilmore1974}. Here, $\boldsymbol{a}_j = (\boldsymbol{R}_j + i \boldsymbol{S}_j)/\sqrt{2}$ and $\boldsymbol{a}_j^{\dagger} = (\boldsymbol{R}_j - i \boldsymbol{S}_j)/\sqrt{2}$ are local creation and annihilation operators, respectively, while $\alpha_j = (r_j + i s_j)/\sqrt{2}$ denotes the complex-valued phase.

As coherent states resolve the identity $\boldsymbol{\mathds{1}}_j = \int \frac{\mathrm{d} r \, \mathrm{d} s}{2 \pi} \, \ket{\boldsymbol{\alpha}_j} \bra{\boldsymbol{\alpha}_j}$, one can associate a positive operator-valued measure (POVM) to pure coherent state projectors, realizable through heterodyne measurements, leading to the bipartite Husimi $Q$-distribution \cite{Husimi1940,Schleich2001,Mandel2013,Schupp2022}
\begin{equation}
    Q(r_1, s_1, r_2, s_2) = (\bra{\alpha_1} \otimes \bra{\alpha_2}) \, \boldsymbol{\rho}_{12} \, (\ket{\alpha_1} \otimes \ket{\alpha_2}).
    \label{eq:GlobalHusimiQDefinition}
\end{equation}
Since coherent states overlap, the Husimi $Q$-distribution violates Kolmogorov's third axiom and has to be considered as a quasi-probability distribution covering phase space. Nevertheless, it is bounded ($0 \le Q \le 1$) and normalized in the sense that $\int \frac{\mathrm{d} r_1 \, \mathrm{d} s_1}{2 \pi} \frac{\mathrm{d} r_2 \, \mathrm{d} s_2}{2 \pi} \, Q(r_1,s_1,r_2,s_2) = 1$. 

We study correlations between the two subsystems with general non-local operators~\cite{Einstein1935,Duan2000,Giovannetti2003},
\begin{equation}
    \boldsymbol{R}_{\pm} = a_1 \boldsymbol{R}_1 \pm a_2 \boldsymbol{R}_2, \,\, \boldsymbol{S}_{\pm} = b_1 \boldsymbol{S}_1 \pm b_2 \boldsymbol{S}_2,
    \label{eq:NonLocalOperatorsDefinition}
\end{equation}
with real and non-negative scaling parameters $a_1,a_2,b_1,b_2 \ge 0$ such that $a_1 b_1 = a_2 b_2$, which fulfill the commutation relations
\begin{equation}
    [\boldsymbol{R}_{\pm}, \boldsymbol{S}_{\pm}] = i (a_1 b_1 + a_2 b_2) \boldsymbol{\mathds{1}}, \quad [\boldsymbol{R}_{\pm}, \boldsymbol{S}_{\mp}] = 0,
    \label{eq:NonLocalOperatorsCommutationRelations}
\end{equation}
showing that pairs of operators with equal indices obey bosonic commutation relations up to normalization. By performing a variable transformation in \eqref{eq:GlobalHusimiQDefinition} we obtain
\begin{equation}
    \begin{split}
        &Q (r_+,s_+,r_-,s_-) = \frac{1}{4 a_1 a_2 b_1 b_2} \\
        &\quad\times Q \left(\frac{r_+ + r_-}{2a_1}, \frac{s_+ + s_-}{2b_1}, \frac{r_+ - r_-}{2a_2}, \frac{s_+ - s_-}{2b_2} \right),
    \end{split}
    \label{eq:GlobalHusimiQNonLocal}
\end{equation}
where the prefactor ensures normalization $\int \frac{\mathrm{d} r_+ \, \mathrm{d} s_+}{2 \pi} \frac{\mathrm{d} r_- \, \mathrm{d} s_-}{2 \pi} \, Q (r_+,s_+,r_-,s_-) = 1$. We marginalize the latter over the mixed pairs $(r_-,s_+)$ or $(r_+,s_-)$
\begin{equation}
    Q_{\pm} \equiv Q (r_{\pm}, s_{\mp}) = \int \frac{\mathrm{d} r_{\mp} \, \mathrm{d} s_{\pm}}{2 \pi} \, Q (r_+,s_+,r_-,s_-),
    \label{eq:MarginalHusimiQ}
\end{equation}
resulting in distributions over the variables $(r_{\pm}, s_{\mp})$. We will show in the following that these distributions are non-trivially constrained for all separable states.

\textit{Entanglement criteria} --- A well-known necessary condition for the separability of a given state $\boldsymbol{\rho}_{12}$ is the Peres-Horodecki (PPT) criterion \cite{Peres1996,Horodecki1996,Horodecki2009}. Every separable state has a non-negative partial transpose, i.e. the operator $\boldsymbol{\rho}'_{12}$ obtained from $\boldsymbol{\rho}_{12} \to \boldsymbol{\rho}'_{12} = \left(\boldsymbol{\mathds{1}}_1 \otimes \boldsymbol{T}_2 \right) (\boldsymbol{\rho}_{12})$ is non-negative $\boldsymbol{\rho}'_{12} \ge 0$, where $\boldsymbol{T}_2$ denotes partial transposition in subsystem $2$. Hence, $\boldsymbol{\rho}'_{12}$ is \textit{physical} for all separable states, implying that derived distributions are constrained by the uncertainty principle. In case of the Husimi $Q$-distribution, this is most generally expressed by the Lieb-Solovej theorem \cite{Lieb2014,Schupp2022}. Ultimately being a majorization relation in phase space \footnote{See \cite{PRA} for a detailed discussion on continuous marjorization theory}, it states that
\begin{equation}
    \int \frac{\mathrm{d}r \, \mathrm{d} s}{2 \pi} f (Q) \ge \int \frac{\mathrm{d}r \, \mathrm{d} s}{2 \pi} f (\bar{Q}),
    \label{eq:LiebSolovejTheorem}
\end{equation}
for all concave $f: [0,1] \to \mathbb{R}$ with $f(0)=0$.

In phase space, the partial transpose $\boldsymbol{T}_2$ has a simple geometric representation. It corresponds to a sign change in the momentum type variable $s_2 \to - s_2$ \cite{Simon2000}. Following \eqref{eq:NonLocalOperatorsDefinition} this implies $s_{\pm} \to s_{\mp}$ and hence $Q (r_{\pm}, s_{\pm}) \to Q (r_{\pm}, s_{\mp})$. Then, the PPT criterion implicates that the Lieb-Solovej theorem \eqref{eq:LiebSolovejTheorem} has to be fulfilled with respect to the vacuum in the variable pairs $(r_{\pm}, s_{\pm})$ \textit{after} partial transposition for all separable states. Defining the witness functional
\begin{equation}
    \mathcal{W}_f = \int \frac{\mathrm{d} r_{\pm} \, \mathrm{d} s_{\mp}}{2 \pi} \left[ f (Q_{\pm})  - f \left(\bar{Q}'_{\pm} \right) \right],
    \label{eq:WitnessDefinition}
\end{equation}
where
\begin{equation}
    \bar{Q}'_{\pm} (r_{\pm}, s_{\mp}) = \frac{1}{a_1 b_1 + a_2 b_2} \, e^{- \frac{1}{2} \frac{r^2_{\pm} + s^2_{\mp}}{a_1 b_1 + a_2 b_2}}
    \label{eq:NonLocalVacuumHusimiQ}
\end{equation}
and $f: \mathcal{J} \to \mathbb{R}$ with $\mathcal{J}=[0,\max \{ \max{Q_{\pm}}, (a_1 b_1 + a_2 b_2)^{-1} \}] \subseteq \mathbb{R}^{+}$ is a concave function with $f(0) = 0$ \footnote{Note that only the first term is state-dependent, while the second is fully determined by the function $f$}, allows us to state our main result: $\mathcal{W}_f$ is non-negative for all separable states, i.e.
\begin{equation}
    \boldsymbol{\rho}_{12} \text{ separable} \Rightarrow \mathcal{W}_f \ge 0,
    \label{eq:SeparabilityCriteria}
\end{equation}
for all $f$ under the assumptions stated above. Violation of the latter inequality for any choice of the concave function $f$, the rotation angles $\vartheta_1, \vartheta_2$, or the scaling parameters $a_1,b_1,a_2,b_2$ with $a_1 b_1 = a_2 b_2$, implies that $\boldsymbol{\rho}_{12}$ is entangled. 
In the remainder of this paper, we show advantages offered by the freedom of using any concave function $f$.

\textit{Entropic criteria} --- The condition $\mathcal{W}_f \ge 0$ remains valid when applying a monotonically increasing function $g: \mathbb{R} \to \mathbb{R}$ to both integrals in \eqref{eq:SeparabilityCriteria}, which allows to recover families of entropic criteria. For monomials $f(t) = t^{\beta}$ with $\beta \in (0,1)$ and $g(t) = \frac{1}{1 - \beta} \ln t$ we find
\begin{equation}
    \mathcal{W}_{\beta} = S_{\beta} (Q_{\pm}) - \frac{\ln \beta}{\beta - 1} - \frac{\ln \det \bar{V}'_{\pm}}{2} \ge 0,
    \label{eq:RenyiWehrlCriteria}
\end{equation}
where
\begin{equation}
    S_{\beta} (Q_{\pm}) = \frac{1}{1 - \beta} \ln \left[ \int \frac{\mathrm{d} r_{\pm} \, \mathrm{d} s_{\mp}}{2 \pi} \, Q_{\pm}^{\beta} (r_{\pm},s_{\mp}) \right]
    \label{eq:RenyiWehrlEntropyDefinition}
\end{equation}
denote Rényi-Wehrl entropies \cite{Bengtsson2017} and $\bar{V}'_{\pm} = (a_1 b_1 + a_2 b_2) \mathds{1}$ is the covariance matrix of the vacuum $\bar{Q}'_{\pm}$. The latter extends to $\beta \in (1, \infty)$ as for the convex function $-f$ a monotonically decreasing function $g$ restores the overall non-negativity of the witness. In the limit $\beta \to 1$ \eqref{eq:RenyiWehrlEntropyDefinition} reduces to the ordinary Wehrl entropy \cite{Wehrl1978,Wehrl1979} and the witness functional \eqref{eq:RenyiWehrlCriteria} boils down to the entropic criteria reported in \cite{Haas2022a} for $a_1=b_1=a_2=b_2=1$, a fact which also follows directly from \eqref{eq:SeparabilityCriteria} with $f(t)=-t \ln t$.

\textit{Second moment criteria} --- Starting from the covariance matrix of the marginal $Q_\pm$ of an arbitrary Husimi $Q$-distribution
\begin{equation}
    V_{\pm} = \begin{pmatrix}
        \sigma_{r_{\pm}}^2 + \frac{a_1^2 + a_2^2}{2} & \sigma_{r_{\pm} s_{\mp}} \\
        \sigma_{r_{\pm} s_{\mp}} & \sigma_{s_{\mp}}^2 + \frac{b_1^2 + b_2^2}{2}
          \end{pmatrix},
    \label{eq:CovarianceMatrixDefinition}
\end{equation}
where $\sigma^2_{r_{\pm}}, \sigma^2_{s_{\mp}}$ and $\sigma_{r_{\pm} s_{\mp}}$ denote the second moments of the Wigner $W$-distribution, we find second moment criteria from \eqref{eq:RenyiWehrlCriteria} in the limit $\beta \to 1$, 
\begin{equation}
    \mathcal{W}_{\det V_{\pm}} = \det V_{\pm} - \det \bar{V}'_{\pm} \ge 0,
    \label{eq:SecondMomentCriteria}
\end{equation}
as $S_1 (Q_{\pm}) \le 1 + \frac{1}{2} \ln \det V_{\pm}$ for all $Q_{\pm}$ with fixed $V_{\pm}$. For Gaussian states with a Husimi $Q$-distribution of the form (we consider centralized states w.l.o.g.)
\begin{equation}
    Q_{\pm} (r_{\pm}, s_{\mp}) = \frac{1}{Z} \, e^{-\frac{1}{2} (r_{\pm}, s_{\mp})^T V_{\pm}^{-1} (r_{\pm}, s_{\mp})},
    \label{eq:GaussianHusimiQ}
\end{equation}
where $Z = \det^{1/2} V_{\pm}$ is a normalization factor, the criteria \eqref{eq:SeparabilityCriteria} are \textit{equivalent} to the second moment criteria $\mathcal{W}_{\det V_{\pm}} \ge 0$ for \textit{all} concave $f$ with $f(0)=0$ (see \cite{PRA} for a proof). Therefore, the second moment criteria \eqref{eq:SecondMomentCriteria} are optimal in the sense that no stronger bound on $\det V_{\pm}$ can be implied from \eqref{eq:SeparabilityCriteria} as this would be in contradiction with the latter equivalence in case of Gaussian distributions. Taking $f(t) \neq - t \ln t$ and maximizing the witness functional $\mathcal{W}_f$ over $Q_{\pm}$ for fixed $V_{\pm}$ can only lead to weaker second moment criteria.

\begin{figure}[t!]
    \centering
    \includegraphics[width=0.99\columnwidth]{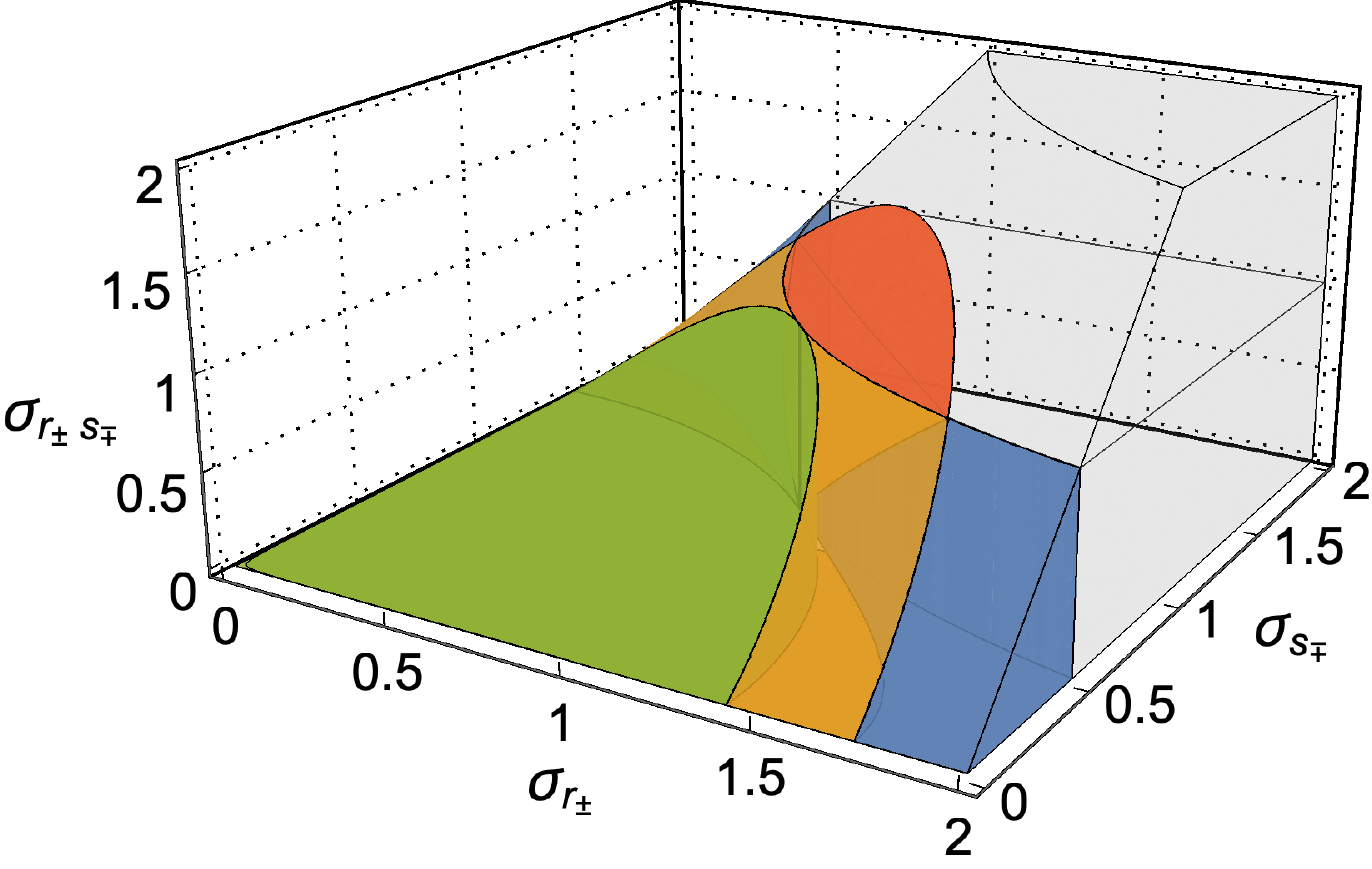}
    \caption{Comparison of witnessed regions of second moment criteria for the general Gaussian distribution \eqref{eq:GaussianHusimiQ} with covariance matrix \eqref{eq:CovarianceMatrixDefinition} and $a_1=b_1=a_2=b_2=1$ together with the allowed region from non-negativity of the Wigner covariance matrix (gray). The green region is detected by all three, but the DGCZ criteria fail beyond. The yellow region is witnessed by both the MGVT and our criteria, while the blue (red) regions are only witnessed by the MGVT (our) criteria, respectively. See also \cite{PRA}.}
    \label{fig:SecondMomentCriteriaCompact}
\end{figure}

We compare \eqref{eq:SecondMomentCriteria} to the DGCZ and MGVT criteria for fixed $a_1, b_1, a_2, b_2$ in \autoref{fig:SecondMomentCriteriaCompact}. Our criteria imply the DGCZ criteria and hence are necessary and sufficient for separability in case of Gaussian states (when optimized over $a_1,a_2,b_1,b_2$). After appropriate optimizations our criteria and the MGVT criteria are equivalent (see \cite{PRA}).

\textit{Example state} --- We benchmark the Rényi-Wehrl criteria \eqref{eq:RenyiWehrlCriteria} against the STW criteria \cite{Saboia2011}, i.e. the strongest entropic criteria based on marginal distributions $f_\pm(r_\pm), g_\mp(s_\mp)$,
\begin{equation}
    \begin{split}
        \mathcal{W}_{\text{STW}} &= S_{\alpha}(f_{\pm}) + S_{\beta}(g_{\mp}) - \frac{\ln \det \bar{V}'_{\pm}}{2} \\
        &+ \frac{1}{2 (1 - \alpha)} \ln \frac{\alpha}{\pi} + \frac{1}{2 (1 - \beta)} \ln \frac{\beta}{\pi},
    \end{split}
    \label{eq:SaboiaCriteria}
\end{equation}
with the marginal Rényi entropies being constrained by $1/\alpha + 1/\beta = 2$. We consider the non-Gaussian state described by the wavefunction \cite{Gomes2009a,Gomes2009b,Nogueira2004,Agarwal2005,Saboia2011,Walborn2009}
\begin{equation}
    \psi (r_1, r_2) = \frac{r_1 + r_2}{\sqrt {\pi \sigma_{-} \sigma^3_{+}}} \, e^{- \frac{1}{4} \left[ \left(\frac{r_1 + r_2}{\sigma_+} \right)^2 + \left(\frac{r_1 - r_2}{\sigma_-}\right)^2 \right]},
    \label{eq:ExampleStateWaveFunction}
\end{equation}
which is entangled for all $\sigma_+, \sigma_- \ge 0$. We choose $a_1 = b_1 = a_2 = b_2 = 1$, and allow for an arbitrary angle $\phi \in [0, 2\pi)$ in the $(r_{\pm}, s_{\mp})$ variables \footnote{Let us stress that our criteria are invariant under such rotations, which can be achieved with suitably chosen $\vartheta_1, \vartheta_2$ (see \cite{PRA}), while for marginal based criteria angle tomography has to be performed} and equal amounts of local squeezing leading effectively to squeezing $\Xi = \text{diag} (\xi, 1/\xi)$ with $\xi > 0$ in the $(r_{\pm}, s_{\mp})$ variables (see \cite{PRA} for explicit expressions).

\begin{figure}[t!]
    \centering
    \includegraphics[width=0.9\columnwidth]{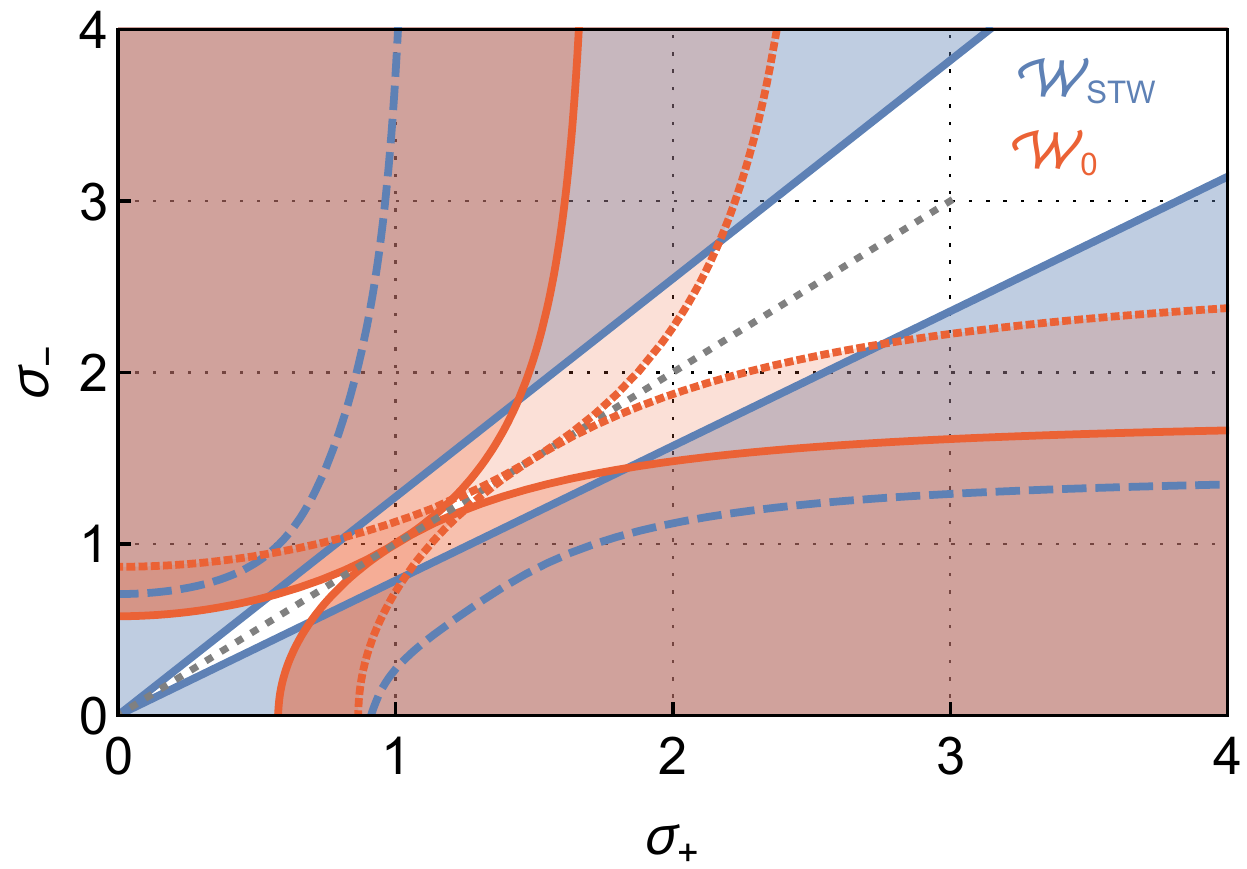}
    \caption{Comparison of witnessed regions of optimal Rényi-Wehrl criteria \eqref{eq:RenyiWehrlCriteria} (red) and optimal STW criteria \eqref{eq:SaboiaCriteria} (blue) for the state \eqref{eq:ExampleStateWaveFunction} with $a_1 = b_1 = a_2 = b_2 = 1$ and $\phi=0, \xi=1$ (straight), $\phi = 0, \xi = 3/2$ (dotted), $\phi = \pi/4, \xi = 1$ (dashed). Our Rényi-Wehrl criteria \eqref{eq:RenyiWehrlCriteria} outperform the STW criteria \eqref{eq:SaboiaCriteria} as after optimizing over $\xi$ we can witness entanglement for all $\sigma_+ \neq \sigma_-$ (gray dashed line). See also \cite{PRA}.}
    \label{fig:MarginalVsPhaseSpaceCriteriaCompact}
\end{figure}

In \autoref{fig:MarginalVsPhaseSpaceCriteriaCompact} we compare the witnessed regions of \eqref{eq:RenyiWehrlCriteria} (red) and \eqref{eq:SaboiaCriteria} (blue) after optimization over the entropic orders in terms of $\sigma_+, \sigma_-$ for three characteristic choices of $\phi$ and $\xi$ reflecting the different invariances of the two sets of criteria. The straight curves correspond to $\phi = 0, \xi = 1$, while the dashed and dotted curves show \eqref{eq:RenyiWehrlCriteria} for $\xi = 3/2$ and \eqref{eq:SaboiaCriteria} for $\phi = \pi/4$, respectively. We observe that after optimizing over $\xi$, our Rényi-Wehrl criteria \eqref{eq:RenyiWehrlCriteria} witness entanglement for all values of $\sigma_+ \neq \sigma_-$ (equality indicated by gray dashed line) in the limit $\beta \to 0$. In contrast, the STW criteria \eqref{eq:SaboiaCriteria}, which witness entanglement only for $\sigma_- / \sigma_+ \ge \pi/4$ and $\sigma_- / \sigma_+ \le 4/\pi$ when $\phi = 0$ and $\alpha \to 1/2$ \cite{Saboia2011}, can only be weakend by improper alignment of the coordinate axes $\phi > 0$ (in which case the optimal $\alpha$ differs in general). 

Instead, for genuinely non-Gaussian states characterized by higher-order correlations or negativities in their Wigner $W$-distributions (for example NOON states), the linear non-local variables \eqref{eq:NonLocalOperatorsDefinition} typically fail to accurately capture entanglement. This applies to all criteria of such type, e.g. \cite{Simon2000,Duan2000,Mancini2002,Giovannetti2003,Walborn2009,Saboia2011}, and thus also to our criteria \eqref{eq:SeparabilityCriteria} (see e.g.\ gray dashed line in \autoref{fig:MarginalVsPhaseSpaceCriteriaCompact} representing the first NOON state).

\textit{Discretized distributions} --- We incorporate the effect of finite resolution, which is relevant for coarse-grained measurements \cite{Rudnicki2012,Tasca2013,Landon2018,Schneeloch2018,Schneeloch2019,PRA}, by discretizing phase space into rectangular tiles $\delta_{j k}$ according to
\begin{equation}
    r^j_{\pm} = j \, \delta r_{\pm}, \quad s^k_{\mp} = k \, \delta s_{\mp},
\end{equation}
where $j,k \in \mathds{Z}$ label the discrete coordinates ($r_{\pm}^j, s_{\mp}^k$), such that $\Delta = \delta r_{\pm} \delta s_{\mp}/(2 \pi)$ denotes the phase space area element of the $(j,k)$-th tile $\delta_{j k}$ centered at ($r_{\pm}^j, s_{\mp}^k$) and hence $\delta_{jk} = [\delta r_{\pm} (j - 1/2), \delta r_{\pm} (j + 1/2)] \times [\delta s_{\mp} (k - 1/2), \delta s_{\mp} (k + 1/2)]$. The resolution factors $\delta r_{\pm}, \delta_{s_{\mp}}$ are proportional to their local analogs, i.e. the resolutions of the local detectors. The probability of observing an event within the $(j,k)$-th tile is given by
\begin{equation}
    Q^{j k}_{\pm} = \int_{\delta_{jk}} \frac{\mathrm{d} r_{\pm} \, \mathrm{d} s_{\mp}}{2 \pi} \, Q_{\pm}(r_{\pm}, s_{\mp}),
    \label{eq:DiscretizedHusimiQHistogramDefinition}
\end{equation}
which is a discrete quasi-probability distribution normalized to unity in the sense that $\sum_{j,k=-\infty}^{\infty} Q^{j k}_{\pm} = 1$.

\begin{figure}[t!]	
    \centering
    \includegraphics[width=0.97\columnwidth]{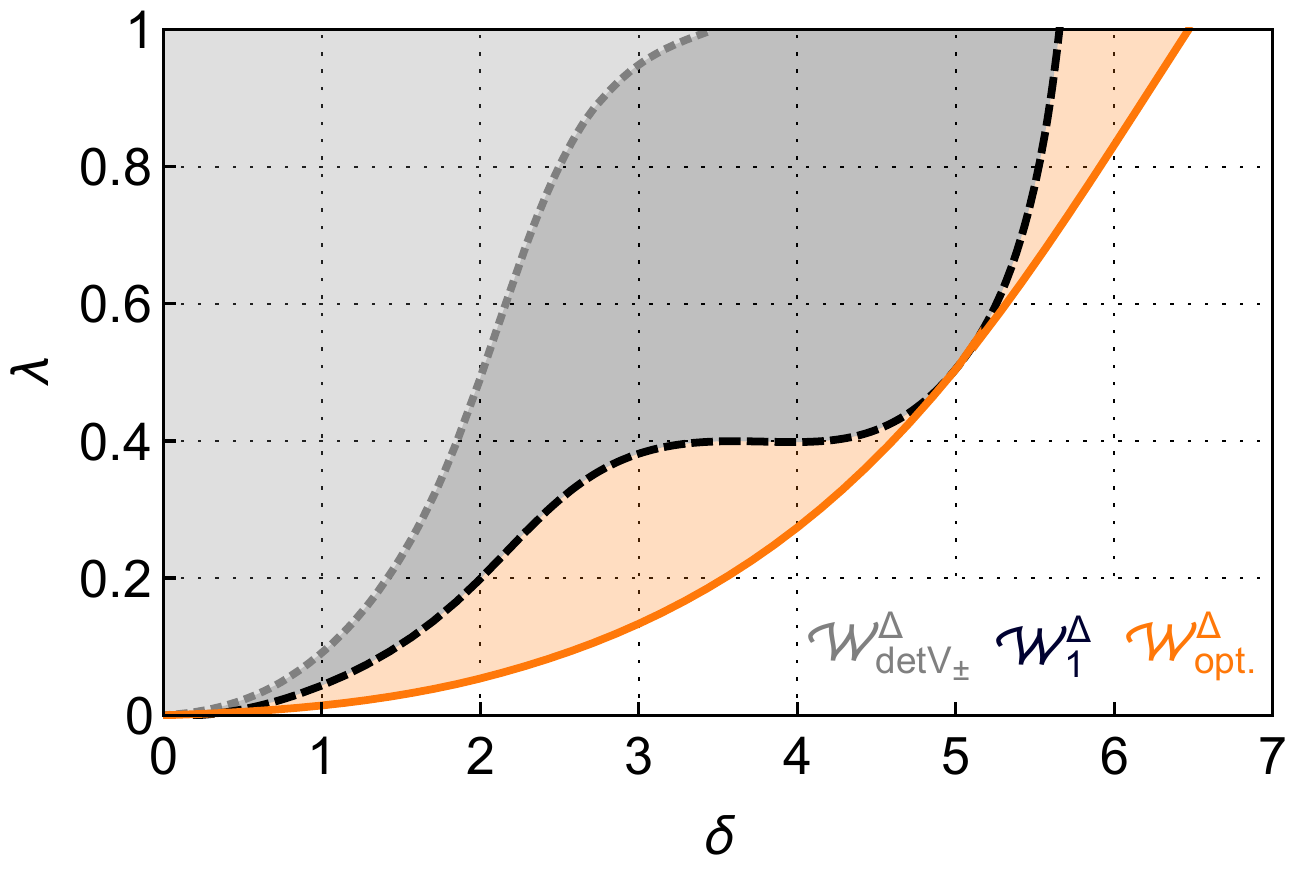}
    \caption{Witnessed regions of discretized second moment (gray dotted), Wehrl entropic (black dashed) and optimal Rényi-Wehrl entropic (orange solid) criteria for a TMSV state of given squeezing $\lambda$ as a function of the grid spacing $\delta$.  Optimization over $\beta$ enables entanglement certification for much larger tilings. See also \cite{PRA}.}
    \label{fig:DiscretizedCriteria}
\end{figure}

Discrete measures of localization such as variance and entropies can underestimate their true continuous values, potentially leading to false-positive demonstrations of entanglement if taken as estimates for their continuous analogs \footnote{Consider the extreme case where all measured data points lie within a single tile s.t. all measures of localization evaluate to zero}. Therefore, uncertainty relations and entanglement criteria are formulated for discrete approximations of the true continuous distributions \cite{Rudnicki2012,Tasca2013,Coles2017}. The distribution $Q_{\pm} (r_{\pm}, s_{\mp})$ is approximated by the density of \eqref{eq:DiscretizedHusimiQHistogramDefinition} over every tile
\begin{equation}
        \hspace{-0.15cm}Q^{\Delta}_{\pm} \equiv Q^{\Delta}_{\pm} (r_{\pm}, s_{\mp}) = \sum_{j,k=-\infty}^{\infty}
        \begin{cases}
             \frac{Q^{j k}_{\pm}}{\Delta} &(r_{\pm}, s_{\mp}) \in \delta_{j k}, \\
            0 &\text{else},
        \end{cases}
    \label{eq:DiscretizedHusimiQDefinition}
\end{equation}
such that $\int \frac{\mathrm{d}r_{\pm} \, \mathrm{d}s_{\mp}}{2 \pi} \, Q^{\Delta}_{\pm} (r_{\pm}, s_{\mp}) = 1$ and $Q^{\Delta}_{\pm} (r_{\pm}, s_{\mp}) \to Q_{\pm} (r_{\pm}, s_{\mp})$ in the continuum limit $\Delta \to 0$. In \cite{PRA} we prove the inequality
\begin{equation}
    \mathcal{W}_f \le \mathcal{W}^{\Delta}_f = \int \frac{\mathrm{d} r_{\pm} \, \mathrm{d} s_{\mp}}{2 \pi} \left[ f (Q^{\Delta}_{\pm})  - f \left(\bar{Q}'_{\pm} \right) \right],
    \label{eq:DiscretizedWitnessDefinition}
\end{equation}
implying that all separable states fulfill (weaker) criteria $\mathcal{W}^{\Delta}_f \ge 0$ for coarse-grained measurements. As $\mathcal{W}_f$ and $\mathcal{W}^{\Delta}_f$ are of the same form and $Q^{\Delta}_{\pm}$ is a quasi-probability density function, the arguments leading to entropic \eqref{eq:RenyiWehrlCriteria} and second moment criteria \eqref{eq:SecondMomentCriteria} hold for $Q^{\Delta}_{\pm}$ as well, giving us their analogs for the continuous approximation $Q^{\Delta}_{\pm}$. Note that all these findings can be generalized to arbitrary discretization schemes (see \cite{PRA}).

\begin{figure}[t!]	
    \centering
    \includegraphics[width=0.99\columnwidth]{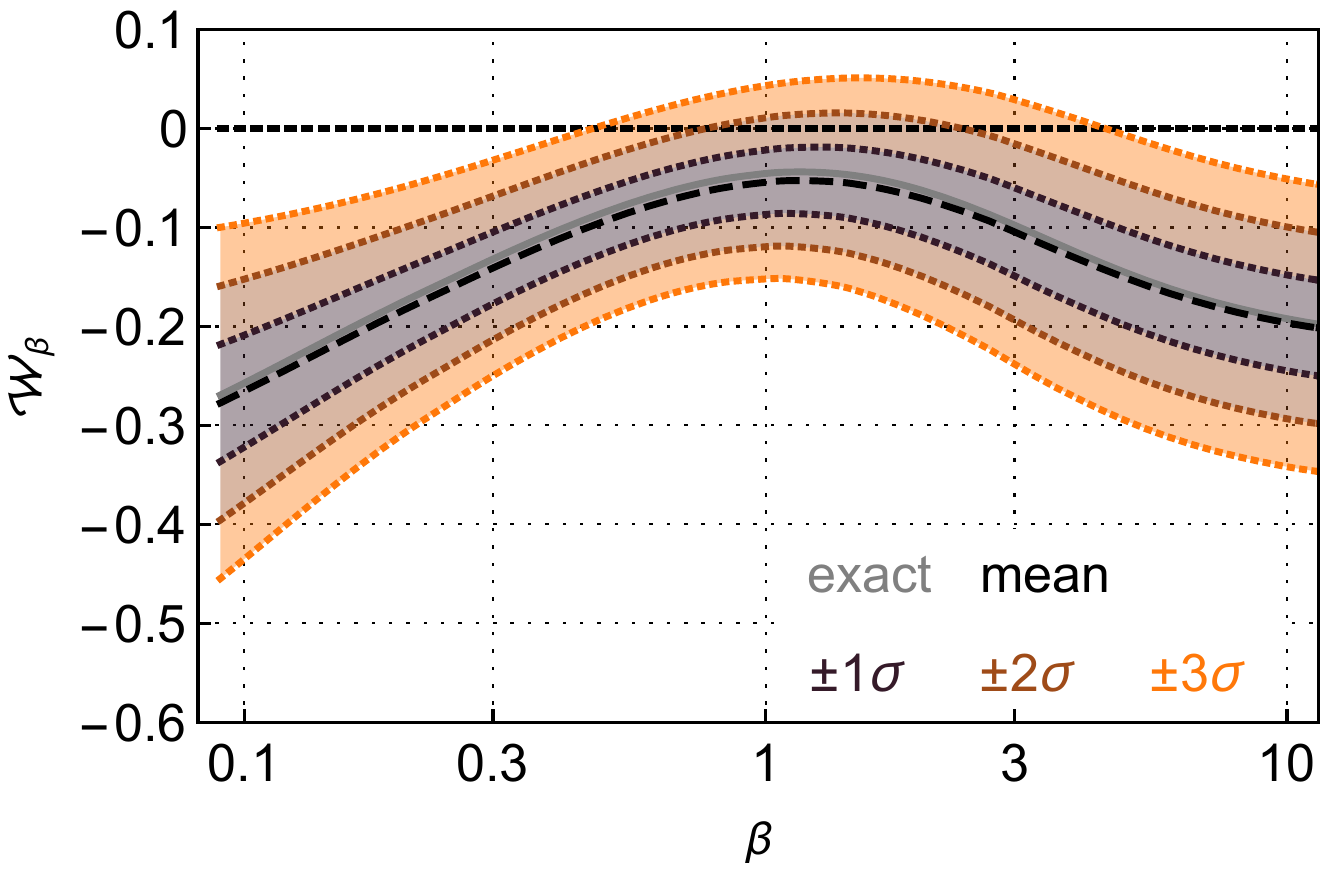}
    \caption{Mean and confidence intervals of Rényi-Wehrl criteria \eqref{eq:RenyiWehrlCriteria} over $\beta$ for $10^3$ samples estimated from $10^2$ repititions for a mixture of TMSV states. For large $\beta \approx 10$ the tails of the estimated distribution are suppressed and entanglement is certified with large confidence. See also \cite{PRA}.}
    \label{fig:SampledCriteria}
\end{figure}

All discretized criteria do never return false positives, but witnessing entanglement is hindered for larger tilings. Optimizing over $f$ substantially improves the detection capabilities, which we illustrate in \autoref{fig:DiscretizedCriteria} for the experimentally relevant case of a two-mode squeezed vacuum (TMSV) state and fixed quadratic tiles $\delta \equiv \delta r_{\pm}^j = \delta s_{\mp}^k$ with $\delta$ being proportional to the local detector resolution, by plotting the witnessed regions of the second moment (gray dotted), Wehrl entropic (black dashed) and the optimal Rényi-Wehrl entropic (orange solid) criteria for every squeezing $\lambda \in [0,1]$ as a function of the grid spacing $\delta$. Although the underlying distribution is of Gaussian form, the optimization over $\beta$ greatly enlarges the detected region especially for small squeezings ($\lambda \approx 0.2$), which is the regime where entanglement detection is typically most challenging in experiments.

\textit{Sampled distributions} --- Finally, we consider the experimentally relevant scenario where entanglement is to be certified on the basis of a small number of experimental samples drawn from the Husimi $Q$-distribution \cite{Kunkel2019,Kunkel2021}. 
The goal here is to maximize statistical significance, i.e. the signal-to-noise ratio with which entanglement is detected. The estimation of functionals of probability density functions from samples, needed for evaluating our entanglement criteria \eqref{eq:SeparabilityCriteria}, is a challenging task in statistical data analysis. It generally requires density estimation of the underlying distribution, while low moments can be estimated directly from the data. Nevertheless, for non-Gaussian states undetectable by second moment criteria (generalized) entropic criteria may be the only way of certifying their entanglement. We show that a suitable choice of the function $f$ in our criteria can lead to an increased signal-to-noise ratio.

We consider $10^3$ samples from a weighted mixture of two displaced TMSV states with Husimi $Q$-distribution
\begin{equation}
    \begin{split}
        Q_{\pm} (r_{\pm}, s_{\mp}) = (1-p) & \, \frac{1+\lambda}{2} \, e^{- \frac{1 + \lambda}{4} \left[ (r_{\pm} - r)^2 + s^2_{\mp} \right]} \\
        +\, p & \, \frac{1+\lambda}{2} \, e^{- \frac{1 + \lambda}{4} \left[ (r_{\pm} + r)^2 + s^2_{\mp} \right]},
        \label{eq:HusimiQTMSVMixture}
    \end{split}
\end{equation}
for opposite displacements $r = \pm 2$, equal squeezings $\lambda = 0.8$ and an unbalanced weight $p=0.3$ \footnote{The second moment criteria \eqref{eq:SecondMomentCriteria} do not witness entanglement at all in this case.}. We estimate the distribution using a Gaussian mixture model (see \cite{PRA} for details), evaluate the family of Rényi-Wehrl criteria \eqref{eq:RenyiWehrlCriteria} for a range of $\beta$-values with $a_1 = b_1 = a_2 = b_2 = 1$ and repeat this procedure $10^2$ times in order to estimate confidence intervals. The results are shown in \autoref{fig:SampledCriteria}, indicating that large $\beta \approx 10$ allows for entanglement certification within significantly larger confidence intervals compared to the standard entropic criteria $\beta \to 1$. 
The reason for this improvement is that for large $\beta$ the influence of regions of small probability, which incur the largest statistical error due to sparse samples, is suppressed.
Surprisingly, also $\beta\ll 1$ yields a reduced statistical error. We attribute this to the prior knowledge about the functional form of the tails of the distribution that is built in to the employed Gaussian mixture model.

\textit{Conclusions and Outlook} --- We have shown that our Husimi-$Q$ based entanglement criteria outperform marginal based criteria in many respects. We emphasize that the Husimi $Q$-distribution completely characterizes a quantum state, just like any other phase space representation, and thus entanglement detection amounts to the task of efficient \emph{extraction} of the relevant information from the respective measured data distributions. Here, our criteria excel by offering vast opportunities of optimal classical post-processing. Our approach of using coherent state projections for entanglement detection opens up a path towards deriving general entanglement criteria for other physically relevant systems such as quantum spins.

\textit{Acknowledgements} --- We thank Oliver Stockdale, Stefan Floerchinger and Markus Oberthaler for discussions on the subject and Célia Griffet for comments on the manuscript. M.G. is supported by the Deutsche Forschungsgemeinschaft (DFG, German Research Foundation) under Germany's Excellence Strategy EXC 2181/1 - 390900948 (the Heidelberg STRUCTURES Excellence Cluster) and under SFB 1225 ISOQUANT - 273811115. T.H. acknowledges support by the European Union under project ShoQC within ERA-NET Cofund in Quantum Technologies (QuantERA) program.


\bibliography{references.bib}

\end{document}